\documentstyle[12pt]{article}

\newcommand{\EQ}{\begin{equation}}
\newcommand{\EN}{\end{equation}}

\begin{document}

\topmargin 0pt
\oddsidemargin 5mm
\newcommand{\NP}[1]{Nucl.\ Phys.\ {\bf #1}}
\newcommand{\PL}[1]{Phys.\ Lett.\ {\bf #1}}
\newcommand{\NC}[1]{Nuovo Cimento {\bf #1}}
\newcommand{\CMP}[1]{Comm.\ Math.\ Phys.\ {\bf #1}}
\newcommand{\PR}[1]{Phys.\ Rev.\ {\bf #1}}
\newcommand{\PRL}[1]{Phys.\ Rev.\ Lett.\ {\bf #1}}
\newcommand{\MPL}[1]{Mod.\ Phys.\ Lett.\ {\bf #1}}
\newcommand{\JETP}[1]{Sov.\ Phys.\ JETP {\bf #1}}
\newcommand{\TMP}[1]{Teor.\ Mat.\ Fiz.\ {\bf #1}}

\renewcommand{\thefootnote}{\fnsymbol{footnote}}

\newpage
\setcounter{page}{0}
\begin{titlepage}
\begin{flushright}
SISSA-EP-72
\end{flushright}
\vspace{0.5cm}
\begin{center}
{\large  Renormalization group trajectories from resonance factorized
S-matrices} \\
\vspace{1cm}
\vspace{1cm}
{\large  M\'arcio Jos\'e  Martins
\footnote{on leave from Departamento de Fisica, Universidade Federal de
S.Carlos, C.P. 676 - S.Carlos 13560, Brazil}
\footnote{martins@itssissa.bitnet}} \\
\vspace{1cm}
{\em International School for Advanced Studies\\
34014, Strada Costiera 11, Trieste,
 Italy}\\
\end{center}
\vspace{1.2cm}

\begin{abstract}
We propose and investigate a large class of models possessing resonance
factorized S-matrices. The associated Casimir energy describes a rich
pattern of renormalization group trajectories related to flows in the
coset models based on the simply laced Lie Algebras. From a simplest
resonance S-matrix, satisfying the ``$\phi^3$-property'', we predict new
flows in non-unitary minimal models.
\end{abstract}
\vspace{.5cm}
\centerline{Pacs numbers: 05.50.+q, 64.60.Cn, 75.10.Jm, 75.10.Hk}
\vspace{.5cm}
\centerline{March 1992}
\vspace{.3cm}
\end{titlepage}

\renewcommand{\thefootnote}{\arabic{footnote}}
\setcounter{footnote}{0}

\newpage
Recently Al.Zamolodchikov has introduced the ``staircase model'' \cite{AL1},
which is defined by a S-matrix of a single massive particle m with amplitude
$S(\theta,\theta_0)=(\sinh(\theta)-i \cosh(\theta_0))/
(\sinh(\theta)+i \cosh(\theta_0))$. The important characteristic of this
S-matrix is to possess two resonance poles at $\theta= -\frac{i \pi}{2} \pm
\theta_0$. The ultraviolet behaviour is governed by a theory of central
charge $c=1$. At intermediate distances, however, this model exhibits a rather
rich renormalization group~(RG) trajectories by varying the real parameter
$\theta_0$. For $\theta_0$ large enough, Al.Zamolodchikov \cite{AL1} has
found that the associated Casimir energy~(defined in a torus of radius R)
interpolates between the central charges $c_p=1-6/p(p+1)$,$p=3,4,...$ of the
minimal models $M_p$. Roughly, the Casimir energy $E_0(R,\theta_0)$ forms
plateaux of approximately length $\frac{\theta_0}{2}$ in each value of $c_p$
before smoothly crossing over to the next lower fixed point $c_{p-1}$. In the
literature
the flow, $M_p \rightarrow M_{p-1}$, is also known \cite{A,CA} as the effect of
the
$\phi_{1,3}$ perturbation to the critical point $M_p$.

The purpose of this Letter is to extend these ideas to  more general
classes of flows. We propose a resonant $Z(N)$-factorizable scattering theory
in which the behaviour of its associated
RG trajectories will be related to a certain deformation of the
minimal $W(A_{N-1})$ conformal models. It turns out that this theory is
related to the scattering of the $A_{N-1}$ Toda model \cite{AFZ} for complex
values of its coupling constant. This relation allows us to easily conjecture
the resonance scattering of the D and E Lie algebras. This last
relation is only formal, due to the fact that the Toda Lagrangian needs an
extra meaning for complex values of its coupling constant. However,
the associated
scattering theory is perfectly
well defined, producing typical RG trajectories of
flows in the deformed coset models based on simply laced Lie groups.
Finally, new flows
are predicted in non-unitary minimal models from the simplest resonance
scattering
theories satisfying the ``$\phi^3$-property''.

We first start by describing the resonance $Z(N)$ scattering. The spectrum
consists of a set of particles and antiparticles with masses
$m_i=\sin(\frac{i \pi}{N})/ \sin(\frac{\pi}{N}),i=1,2,...,N-1$\cite{KW}. The
antiparticle appears in the particle-particle amplitude, and the
factorizability
implies \cite{KUW} that the only constrains are the
crossing and the unitarity conditions,
\EQ
S_{i,j}(\theta) S_{i,j}(-\theta)=1,~~ S_{i,j}(\theta)=S_{N-j,i}(i \pi -\theta)
\EN
where $N-j$ is the antiparticle of the $j$-particle.

There is a family of solutions for Eq.(1) as a function of a real parameter
$\theta_0$, which is responsible for the resonance poles. The minimal solution
reads,
\EQ
S_{1,1}(\theta,\theta_0)=\frac{\sinh\frac{1}{2}(\theta+i\frac{2 \pi}{N})}
{\sinh\frac{1}{2}(\theta-i\frac{2 \pi}{N})}
\frac{\sinh\frac{1}{2}(\theta-\theta_0-i\frac{\pi}{N})}
{\sinh\frac{1}{2}(\theta-\theta_0+i\frac{\pi}{N})}
\frac{\sinh\frac{1}{2}(\theta+\theta_0-i\frac{\pi}{N})}
{\sinh\frac{1}{2}(\theta+\theta_0+i\frac{\pi}{N})}
\EN

The physical pole is at $\theta=\frac{2 \pi i}{N}$, while the resonance poles
appear in the unphysical sheet at
$\theta=-\frac{i \pi}{N} \pm \theta_0$. The other amplitudes $S_{i,j}$ are
obtained from $S_{1,1}$ by applying the bootstrap approach at $\theta=
\frac{i \pi}{N}(i-j +2(a-b))$; a=1,2,...,j-1, b=1,2,...,i-1. For $N=2$,
we obtain Al.Zamolodchikov's model \cite{AL1}.

Here our interest is to study this theory at intermediate distances, by
analyzing the finite volume effects to the Casimir energy. An effective
way to study the Casimir energy $E(R,\theta_0)$ in a geometry of finite
volume R is via the thermodynamic Bethe ansatz~(TBA) approach \cite{YY,ZA2}
at temperature $T=\frac{1}{R}$. For the sake of simplicity let us first
concentrate in the case $N=3$. In this case the Casimir energy $E(R,\theta_0)$
is given by,
\EQ
E(R,\theta_0) = \frac{-m}{\pi}
\int_{-\infty}^{\infty} d\theta cosh(\theta) L(\epsilon)
\EN
where $L(\epsilon)=ln(1+e^{-\epsilon(\theta)})$, $m$ is the mass of the
particle and its antiparticle, and the pseudoenergy $\epsilon(\theta)$
satisfies the following integral equation,
\EQ
\epsilon(\theta) + \frac{1}{2\pi}
\int_{-\infty}^{\infty} d\theta'
\psi(\theta-\theta',\theta_0) L(\epsilon) =
m R \cosh(\theta)
\EN
where $\psi(\theta,\theta_0)=-i\frac{d}{d \theta} \log \left [
(S_{1,1}(\theta,\theta_0)
S_{1,1}(i \pi-\theta,\theta_0)) \right ]$

The ultraviolet limit of the Casimir energy
, $R \rightarrow 0$, is independent of $\theta_0$ and we find the behaviour
$E(R,\theta_0) \simeq -\frac{2 \pi}{6 R}$, which implies that the background
conformal theory has central charge $c=2$ \cite{BCI}. In order to analyze
the behaviour of the function $c(R,\theta_0)=-\frac{6 R E(R,\theta_0)}{\pi}$
for finite values of R we numerically solved Eq.(2), in the convenient
variable $X=\log(mR/2)$, by standard interactive procedure. For $\theta_0=0$,
$c(R,0)$ behaves as a smooth function  between the
ultraviolet~(c=2) and the infrared~(c=0) regimes. By increasing $\theta_0$,
however, we observe that certain plateaux start to form precisely around
the values that parametrize the central charge of the minimal model of
the $W(A_2)$ algebra, namely $c=2(1-12/p(p+1))$, p=4,5,... .For example
at $\theta_0=40$, we notice at least 8 plateaux starting at p=12 and
subsequently visiting the other fixed points p=11,...,4, until
finally reaching the infrared region. In Fig.1(a,b) we show this behaviour
for $\theta_0=20,40$. The same pattern can be viewed from
the beta function along the RG trajectories. Following Al.Zamolodchikov's
notation \cite{AL1}, one can define the beta function as,
\EQ
\beta(g)=-\frac{d}{dX}c(R,\theta_0),~~~~g=2-c(R,\theta_0)
\EN
In Fig.2(a,b), we show $\beta(g)$ for $\theta_0=20,40$. The zeros of
$\beta(g)$ are formed precisely at $g=24/p(p+1)$,p=4,5,..., in accordance
with the plateaux mentioned above.

In the case of general $N$ we should expect a similar behaviour. From
Al.Zamolodchikov's discussion of N=2 and our present results we conclude that
each time that $X \simeq -(p-N)\frac{\theta_0}{2}$ the function $c(R,\theta_0)$
will crossover its value of $c_p=(N-1) \left ( 1-N(N+1)/p(p+1) \right)$,
p=N+1,N+2,..., to the next fixed point with central charge $c_{p+1}$. Indeed,
by linearizing the TBA equations around $X \simeq -(p-N)\frac{\theta_0}{2}$ one
remains with the same equation that describes the flow in the $W(A_{N-1})$
minimal models perturbed by the least $Z(N)$-invariant operator \cite{VMZ}.
However we stress that the bulk of each plateau has the approximately length of
$\frac{\theta_0}{N}$, in agreement with the fact that the finite-size
corrections are N-dependent.

As an important remark we mention that our proposed resonance Z(N)-factorized
model is easily connected to the one of the $A_{N-1}$ Toda field theory
\cite{AFZ}, by making an analytical continuation to the complex
values of the Toda coupling constant. The coupling constant $\alpha$
enters in the S-matrices through a function $b_{Toda}(\alpha)$ \cite{AFZ}.
By setting
$b_{Toda}(\alpha)=\frac{\pi}{N} \pm i \theta_0$ in Eq.(2) we recover the
minimal
S-matrix of the $A_{N-1}$ Toda model \cite{AFZ}. This lead us
to conjecture that
the resonance scattering theories, based on the simple laced Lie algebras ADE,
can be obtained from the corresponding Toda S-matrices \cite{Tudo}. The
resonance parameter $\theta_0$ is introduced
through the simple relation $b_{Toda}(\alpha)=
\frac{\pi}{h} \pm i \theta_0$, where h is the dual Coxeter number of the
respective ADE Lie algebra. Here we have also analyzed the TBA equations
around $X \simeq -(p-h) \frac{\theta_0}{2}$ and performed numerical checks.
Our conclusion is that the ADE
resonance
scattering will produce RG trajectories associated with the flows in
the coset model $G_{p-h} \otimes G_{1}/G_{p-h+1}$~(G=A,D,E)\cite{GKO}
perturbed by
the field $\Phi$ with conformal dimension
$\Delta_{\Phi}=1-h/(p+1)$\cite{VMZ}. The
scaling corrections in the infrared regime is made by the ``dual'' operator
$\tilde{\Phi}$ with conformal dimension $\Delta_{\tilde{\Phi}}=1+h/(p-1)$(for
p=h+1, this field is replaced by the spinless combination
$T \bar{T}$ of the stress energy tensor $T$). The field $\Phi$($
\tilde{\Phi})$ is the analogue of the operators $\phi_{1,3}$($\phi_{3,1}$)
of the minimal models. It has been argued \cite{AL1,MI} that the combination
$\lambda \phi_{1,3} +\tilde{\lambda} \phi_{3,1}$ plays a fundamental role
in the description of Al. Zamolodchikov's staircase model as a perturbed
conformal field theory. In our case, the straightforward generalization will
consider the combination $\lambda \Phi +\tilde{\lambda} \tilde{\Phi}$. We
remark, however, that this picture has to be checked by a careful
analysis of the
finite size corrections to the fixed point \cite{MM}.

Let's us now introduce a resonance scattering model possessing
the ``$\phi^3$-property'' that will be connected with new flows in non-unitary
minimal models with $c<1$. The model consists of a single particle $a$ and
its two-body S-matrix is given by,
\EQ
S_{a,a}(\theta,\theta_0)=\frac{\tanh\frac{1}{2}(\theta+i\frac{2 \pi}{3})}
{\tanh\frac{1}{2}(\theta-i\frac{2 \pi}{3})}
\frac{\tanh\frac{1}{2}(\theta-\theta_0-i\frac{\pi}{3})}
{\tanh\frac{1}{2}(\theta-\theta_0+i\frac{\pi}{3})}
\frac{\tanh\frac{1}{2}(\theta+\theta_0-i\frac{\pi}{3})}
{\tanh\frac{1}{2}(\theta+\theta_0+i\frac{\pi}{3})}
\EN

The pole at $\theta=\frac{2 \pi i}{3}$ produces the
particle itself~($\phi^3$-property) and the resonance poles are located at
$\theta_0=\frac{i \pi}{3} \pm  \theta_0$. It turns out that the amplitude
$S_{a,a}(\theta,\theta_0)$
satisfies the relation $S_{a,a}(\theta,\theta_0)=S_{1,1}(\theta,\theta_0)
S_{1,2}(\theta,\theta_0)$, where $S_{i,j}(\theta,\theta_0)$ are the S-matrices
of the resonance Z(3)-model. The equivalent Toda theory \cite{AFZ} is one
proposed by Milkhailov \cite{MS} as a particular reduction of the Z(3) Toda
model.
{}From the TBA point of view, this implies that
the Casimir energy associated to $S_{a,a}(\theta,\theta_0)$
is precisely half of that of the Z(3)-model. Hence, now the plateaux will be
formed around the values $c_p=1-12/p(p+1),p=4,5...$. This result suggests that
we are dealing with  RG trajectories of the non-unitary minimal models. We
recall
that in this case the Casimir energy is identified with the effective central
charge $c_{eff}=(24 \Delta-c)$, where $\Delta$ is the
lowest conformal dimension \cite{ZU}. Indeed, this is satisfied by the
following
classes of non-unitary minimal models,
\EQ
M_{\frac{q}{2q+1}},~M_{\frac{q+1}{2q+1}},~~q=p-2=2,3,...
\EN

The flow pattern  is well showed in the Fig.(3). We
see that while in the $M_{\frac{q}{2q+1}}$ model the relevant(irrelevant)
operator
associated with the perturbation(infrared corrections)
is $\phi_{1,5}$($\phi_{2,1}$), in the $M_{\frac{q+1}{2q+1}}$ theory the
situation is replaced by $\phi_{2,1}$($\phi_{1,5}$).
In the RG trajectories
the field $\phi_{1,5}$ and $\phi_{2,1}$ interchange their roles of relevant
and irrelevant operators, building an extremely interesting pattern of flow.
For an extra support we also have numerically studied the spectrum of the
simplest case, namely $M_{3/5} +\phi_{2,1} \rightarrow M_{2/5}$. Our results
are compatible with the typical behaviour expected of the RG flows.
Also, in the case of $M_{3/5}$, the field $\phi_{1,5}$(not present on its
Kac-table) is substituted by the level 2 descendent of the $\phi_{1,3}$
operator.

In summary we have discussed rich classes of renormalization RG trajectories
obtained from resonance scattering models based on the ADE Lie algebras. New
flows in non-unitary minimal models have also
been predicted. In our conclusions the
main bulk of technical details were omitted, and they will be presented in a
forthcoming publication \cite{MM}.

\section*{Acknowledgements}

\newpage
\centerline{\bf Figure Captions}
\vspace{0.5cm}
Fig. 1(a,b) The scaling function $c(R,\theta_0)$ for (a)$\theta_0=20$,
and for (b)
$\theta_0=40$
\vspace{0.1cm}\\
Fig. 2(a,b) The beta function $\beta(g)$ for (a) $\theta_0=20$
and for (b) $\theta_0=40$. For $\theta_0=40$ we have omitted the first zero at
$p=5$ in order to better show the remaining zeros of $\beta(g)$
\vspace{0.1cm}\\
Fig. 3 The flow pattern in the non-unitary minimal models
$M_{\frac{q}{2q+1}}; M_{\frac{q+1}{2q+1}},~~q=2,3,...$. The
horizontal(vertical)
arrows represent the relevant(irrelevant) operators defining the
ultraviolet(infrared) corrections to the fixed point.

\begin{thebibliography}{99}
\bibitem{AL1} Al.B. Zamolodchikov, {\em Paris preprint (1991) ENS-LPS-335}
\bibitem{A} A.B. Zamolodchiklov, {\em Yad.Fiz. 46 (1987) 82 [Sov.J.Nucl.Phys.
46 (1987) 1090]}
\bibitem{CA} A.W.W. Ludwig, J.L. Cardy {\em Nucl.Phys.B 285 (1987) 687}
\bibitem{KW} R. Koberle, J.A. Swieca {\em Phys.Lett.B 86 (1979) 209}
\bibitem{KUW} V. Kurak, J.A. Swieca {\em Phys.Lett.B 82 (1979) 289}
\bibitem{YY} C.N. Yang, C.P. Yang, {\em J.Math.Phys.B 10 (1969) 1115}
\bibitem{ZA2} Al.B.Zamolodchikov, {\em Nucl.Phys.B 342 (1990) 695}
\bibitem{BCI} H.W.J.Blote,
J.L.Cardy, M.P.Nightingale, {\em Phys.Rev.Lett.56 (1986) 742}\\
I.Affleck, {\em Phys.Rev.Lett.56 (1986) 746}
\bibitem{VMZ} V.A. Fateev, M.J. Martins, Al.B. Zamolodchikov, unpublished\\
M.J. Martins, {\em Phys.Lett.B 277 (1992) 301}\\
F.Ravanini,{\em Saclay preprint (1992) SPht/92-011}
\bibitem{AFZ} A.E. Arinshtein, V.A. Fateev, A.B. Zamolodchikov {\em Phys.Lett.
B 87(1979) 389}
\bibitem{Tudo} P.Christe, G. Mussardo, {\em Nucl.Phys.B 330 (1990) 465;
Int.J.Mod.Phys. A5 (1990) 4581}\\
H.W.Braden, E.Corrigan, P.E. Dorey, R. Sasaki, {\em Phys.Lett.B 227 (1989)
441; Nucl.Phys.B 338 (1990) 689}\\
C.Destri, H.J.de Vega {\em Phys.Lett.B 233 (1989) 336}
\bibitem{GKO} P. Goddard, A. Kent, D. Olive, {\em Commum.Math.Phys.
103 (1986) 105}
\bibitem{MI} M.Lassig, {\em Julich preprint (1991)}
\bibitem{MS} A.V. Milkhailov, {\em Pisma v ZhETF 30 (1979) 443 [JETP Letters
30 (1979) 414]}
\bibitem{ZU} C.Itzykson, H.Sauler, J.-B.Zuber, {\em Europhys.Lett.2 (1986) 91}
\bibitem{MM} M.J. Martins,(to be published)
\end{thebibliography}
\end{document}